# Evaluation of blackbody radiation shift with temperature associated fractional uncertainty at $10^{-18}$ level for $^{40}$Ca$^+$ ion optical clock

Ping Zhang [1, 2, 3, 4], Jian Cao [1, 2, *], Hua-lin Shu [1, 2], Jin-bo Yuan [1, 2, 3], Jun-juan Shang [1, 2, 3], Kai-feng Cui [1, 2, 3], Si-jia Chao [1, 2, 3], Shao-mao Wang [1, 2, 3], Dao-xin Liu [1, 2, 3], Xue-ren Huang [1, 2, †]

[1] State Key Laboratory of Magnetic Resonance and Atomic and Molecular Physics, Wuhan Institute of Physics and Mathematics, Chinese Academy of Sciences, Wuhan 430071, China

[2] Key Laboratory of Atom Frequency Standards, Wuhan Institute of Physics and Mathematics, Chinese Academy of Sciences, Wuhan 430071, China

[3] University of Chinese Academy of Sciences, Beijing 100049, China

[4] School of Physics and Electrical Engineering, Taizhou University, Taizhou 318000, China

* Corresponding author. Email: caojian@wipm.ac.cn

† Corresponding author. Email: hxueren@wipm.ac.cn

Abstract:

In this paper, blackbody radiation (BBR) temperature rise seen by the $^{40}$Ca$^+$ ion confined in a miniature Paul trap and its uncertainty have been evaluated via finite-element method (FEM) modelling. The FEM model was validated by comparing with thermal camera measurements, which were calibrated by PT1000 resistance thermometer, at several points on a dummy trap. The input modelling parameters were analyzed carefully in detail, and their contributions to the uncertainty of environment temperature were evaluated on the validated FEM model. The result shows that the temperature rise seen by $^{40}$Ca$^+$ ion is 1.72 K with an uncertainty of 0.46 K. It results in a contribution of 2.2 mHz to the systematic uncertainty of $^{40}$Ca$^+$ ion optical clock, corresponding to a fractional uncertainty $5.4 \times 10^{-18}$. This is much smaller than the uncertainty caused by the BBR shift coefficient, which is evaluated to be 4.8 mHz and at $10^{-17}$ level in fractional frequency units.

1 Introduction

Optical clocks, both based on single trapped ions [1-4] and neutral atoms [5-8], have shown great improvements in recent years. Some of them have been confirmed to possess a total systematic uncertainty smaller than Cs microwave frequency standards which are realized as SI second currently [3-6]. While there are a number of systematic shifts should be overcome and one of them is blackbody radiation (BBR) shift which has been one of the largest contributions to the systematic uncertainty for many optical clocks. Especially, for single ion optical clocks which trap ion using RF field, heating is unavoidable as RF absorption in insulators and joule heating in conductors. Thus, it is more difficult to evaluate the temperature precisely for ion optical clocks when compared with the ones using neutral atoms confined in optical lattices.

Optical clock based on $^{40}Ca^+$ ion has been well developed in the past decades due to the advantages on the relatively simple scheme and low cost of clock system [2, 9, 10]. Unlike the one based on $^{27}Al^+$ ion, which is insensitive to BBR [1], $^{40}Ca^+$ ion optical clock possess a large BBR shift which contributes a lot to the clock systematic uncertainty [2]. It has been identified that the uncertainty of BBR shift is basically depended on the uncertainty of the BBR shift coefficient and the uncertainty of temperature measurement on the trap environment [11]. For $^{40}Ca^+$ ion optical clock based on $4S_{1/2}$-$3D_{5/2}$ transition, the first part is the uncertainty of the differential scalar polarizability of $4S_{1/2}$ and $3D_{5/2}$ states, which results in a contribution of fractional frequency uncertainty at $10^{-17}$ level according to the theoretical calculations described in Ref. [11]. The last one is strongly limited by the accuracy of experiment measurement on the environment temperature and also has a contribution at $10^{-17}$ level in fractional frequency units when using temperature sensors [1, 12] or thermal imaging devices [2, 4]. Recently, the trap temperature has been evaluated more precisely via a method based on finite-element method (FEM) modelling [13, 14]. The results indicate that the temperature associated fractional uncertainty of BBR shift has the potential to reach $10^{-18}$ level and will be no longer one of the largest contributions to the systematic uncertainty for most of ion optical clocks.

Here, we evaluate BBR temperature rise seen by $^{40}Ca^+$ ion in a miniature Paul trap via FEM modelling. For validating the FEM model, a dummy trap was established and the temperature of trap electrodes in vacuum chamber measured by thermal camera was calibrated by PT1000 resistance thermometer. The uncertainty of input modelling parameters and their contributions to the uncertainty of the BBR temperature rise were evaluated in detail subsequently. The evaluation result indicates that the temperature rise seen by $^{40}Ca^+$ ion is 1.72 K with an uncertainty of 0.46 K. It causes a contribution of 2.2 mHz to the systematic uncertainty and $5.4 \times 10^{-18}$ in fractional frequency units. Compared to the one caused by the BBR shift coefficient at $10^{-17}$ level, this uncertainty is much smaller and on longer one of the main contributions to the total BBR shift uncertainty of $^{40}Ca^+$ ion optical clock.

2 Trap description and BBR FEM modelling

The miniature Paul trap has been used in $^{40}Ca^+$ ion optical clock study by Wuhan Institute of Physics and Mathematics in China (WIPM) for many years [2, 15, 16]. The improved structure of the miniature Paul trap employed in our system is shown in Figure 1. This trap contains one ring electrode, two endcap electrodes and two compensation electrodes which are all made from molybdenum. The radius of ring electrode wire is 0.25 mm, while the one of endcap electrodes wire as well as potential electrodes wire is 0.5 mm. The ratio of these two radius has been optimized to be 0.5 according to our previous work [15]. All of these electrodes are connected to the titanium wires on titanium support disc. To obtain good electronical and thermal connections, they are fixed together through spot welding. Then, titanium support disc are mounted to a vacuum feedthrough, and both electronical and thermal connections between their electrodes are also achieved through spot welding. The ceramics used in support disc and vacuum feedthrough to insulate the electrodes from them are both alumina.

A FEM model was developed to evaluate the temperature rise of the trap system and the radiation temperature rise seen by the $^{40}Ca^+$ ion. In the model, the Electromagnetic waves, Frequency Domain interface found under the Radio Frequency branch is employed to study the electric field and the Heat Transfer in Solid found under the Heat Transfer branch to study the thermal distribution. RF absorption in insulators and joule heating in conductors are regarded as the two main sources of heat generation, as well as radiation and conduction in solids as the two primary ways of heat removal. A small blackbody sphere with unity emissivity and a thermal conductivity of 1000 W $m^{-1}$ $K^{-1}$ was located at the center of the ring electrode to evaluate the radiation temperature rise there. The diameter of the sphere is 10 um, and the temperature fluctuation on it could be neglected due to its high thermal conductivity. In modelling, a 24.5 MHz RF was added between ring electrode and endcap electrodes according to the one used to trap $^{40}Ca^+$ ion in our optical clock system. The zero-to-peak amplitude of RF is 520 V and 700 V for validating the model and evaluating the temperature rise respectively. In the trap system, besides molybdenum, titanium, and alumina mentioned above, titanium and fused silica were also used at the vacuum chamber and windows. And the main properties of materials used in FEM modelling are listed in table 1.

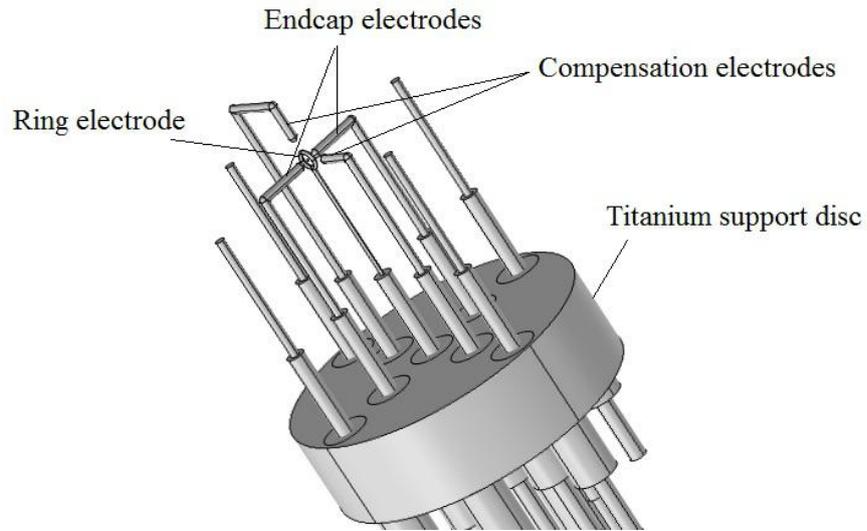

**Figure 1.** Structure of miniature Paul trap. The miniature Paul trap contains one ring electrode, two endcap electrodes and two compensation electrodes which connected to the titanium wires fixed on titanium support disc through spot welding.

**Table 1.** The main properties of materials used in FEM modelling.

| Material | Electrical conductivity [a] (MS m$^{-1}$) | Thermal conductivity (W m$^{-1}$ K$^{-1}$) | Emissivity | Relative permittivity [a] | Dielectric loss tangent [a] |
|---|---|---|---|---|---|
| Molybdenum | 17.6 | 139 | 0.162 | - | - |
| Titanium | 1.82 | 21 | 0.166 | - | - |
| Aluminum | 36.0 | 150 | 0.25 | - | - |
| Alumina | - | 35.4 | 0.9 | 9.5 | 2.0E-4 |
| Fused silica | - | 1.35 | 0.8 | 3.82 | 1.5E-5 |

[a] Electrical conductivity of ceramics, relative permittivity and dielectric loss tangent of metals have not been considered in modelling and assigned by the default value of the finite-element analysis software.
[b] The properties list in this table are main referred to the paper [13] and others which have not list here are given by the finite-element analysis software by default.

3 Results and discussion

3.1 Validation of the FEM model

The FEM model was validated by comparing with the measurements of trap temperature at several points on a dummy trap. Experimentally, it is very difficult to measure the trap temperature accurately by using thermal temperature sensors due to the influence of RF. Thus, a thermal camera (Fluke Ti400) ranging from 7.5 um to 14 um, instead of thermal temperature sensors, has been employed to gauge the trap temperature through CaF window. While, the lower emissivity of molybdenum, as well as lower transmissivity of CaF window in far infrared range, causes the measurement using thermal camera have deviation with the actual temperature. Moreover, a parameter of low emissivity can make the uncertainty of the measurements larger when using thermal camera. To avoid such problems, a calibration of the thermal camera measurements was carried out by comparing with the temperature measured by a calibrated PT1000 resistance thermometer on a same molybdenum plate. For getting the actual temperature, PT1000 resistance thermometer was glued on the molybdenum plate by using silicon heat sink compound. Then the plate together with the thermometer was put into a vacuum chamber to gauge temperature more precisely. For obtaining more data, the plate was heated up by a film heater stuck on the back side at different DC currents. After thermal balance at a fixed heating current, the steady-state temperature of molybdenum was measured by PT1000 resistance thermometer and thermal camera simultaneously. The relationship of these two measurements is exhibited in figure 2. As shown in the figure, the dots are the measured results and the line is obtained by polynomial fitting.

Figure 3 presents the thermal pictures of the miniature Paul trap (a) without an applied RF and (b) with an applied RF. The zero-to-peak amplitude of RF applied on the dummy trap is 520V determined by a calculation described below. Obviously, the trap temperature in figure 3 (b) is higher than the one in figure 3 (a). This means the trap structure has been heated up after adding RF between ring electrode and endcap electrodes. At the surface of the spot welding region, it is brighter than other area. That may be caused by a different reflectivity and emissivity of the spot welding material. By using the software attached with thermal camera, the temperature of ring electrode and endcap electrode is determined to be 296.9 K and 296.8 K respectively. According to the fit line in figure 2, the actual temperature of these two electrodes is estimated to be 300.0 K and 299.6 K. Then, the FEM model was validated by adjusting the structure of insulator which was not acquainted very well and its thermal conductivity until the modelling results coincide with the temperature measurements.

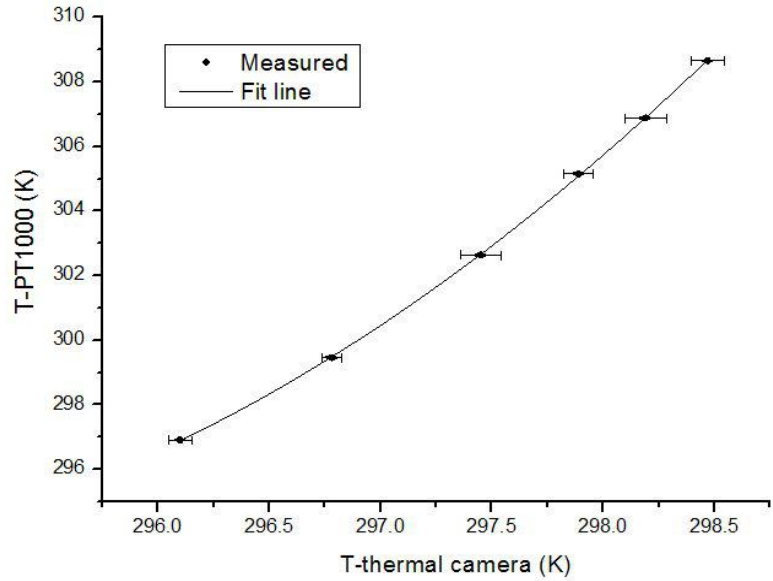

**Figure 2.** The relationship of the molybdenum temperature in the vacuum chamber measured by PT1000 resistance thermometer (Y-axis) and the ones measured by thermal camera (X-axis).

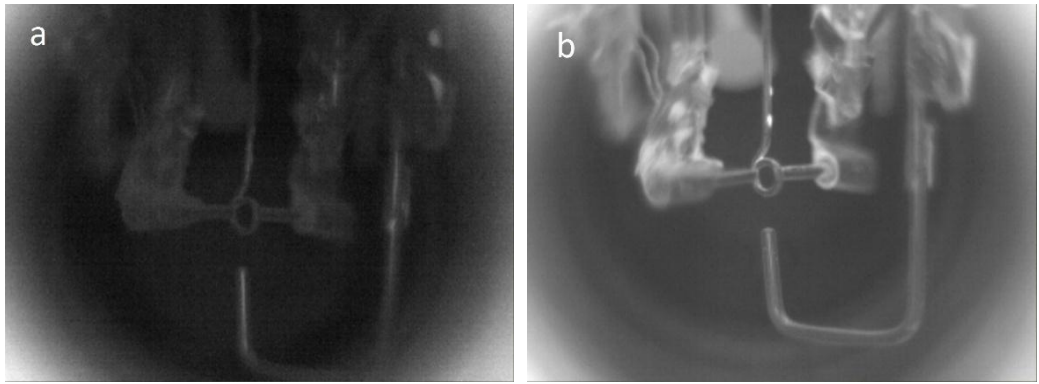

**Figure 3.** Thermal images of the miniature Paul trap (a) without an applied RF and (b) with a 520 V RF at about 24.5 MHz.

3.2 Uncertainty estimation of modelling parameters

For the evaluation of the temperature rise seen by the ion via FEM modelling, the uncertainty is mainly coming from the input modelling parameters such as RF amplitude, dielectric loss tangent of ceramic, thermal conductivity of ceramic, emissivity of trap electrodes and inhomogeneous vacuum chamber temperature [13]. According to the structure of the miniature Paul trap and the opted materials, the main input parameters has been analyzed and estimated conservatively in the following passage. The value of these parameters and their uncertainty used in evaluating are listed in table 2.

Table 2. The value of main input parameters and their uncertainty used in evaluating the uncertainty of radiation temperature rise seen by $^{40}Ca^+$.

| Main Parameter | Value | Uncertainty |
| --- | --- | --- |
| RF amplitude (V) | 700 | 50 |
| Emissivity of Mo | 0.16 | 0.09 |
| Thermal conductivity of alumina (W m$^{-1}$ K$^{-1}$) | 27.5 | 7.5 |
| Dielectric loss tangent of alumina ($10^{-4}$) | 1.5 | 0.5 |
| Initial temperature of electrodes (K) | 292.2 | 1.0 |
| Vacuum chamber temperature (K) | 292.2 | 0.1 |

RF zero-to-peak amplitude of the trap system was measured to be 600 V by using an oscilloscope. While this value has a deviation with the voltage amplitude used to trap $^{40}Ca^+$ ion due to the self-capacity of high-voltage probe in parallel. For determining the amplitude more precisely, following calculation has been carried out. According to Ref. [17], the resonant frequency f of a helical resonator can be calculated from the following equation:

$$f = \frac{1}{2\pi\sqrt{LC}}, \qquad (1)$$

where L is the inductance and C is the capacitance of the trapping system. And the output voltage V of the helical resonator is

$$V = \left(\frac{L}{C}\right)^{1/4} \sqrt{PQ}, \qquad (2)$$

where P is input power and Q is quality factor. For connecting a different load capacitance, we can get the following relation:

$$\frac{V}{V'} = \left(\frac{C'}{C}\right)^{1/4} \frac{\sqrt{Q}}{\sqrt{Q'}} \frac{\sqrt{P}}{\sqrt{P'}} = \left(\frac{f'}{f}\right)^{1/2} \left(\frac{Q}{Q'}\right)^{1/2} \left(\frac{P}{P'}\right)^{1/2}, \qquad (3)$$

where V', C', Q', P', f' are the output voltage, total capacitance, quality factor, input power and resonant frequency respectively at a different load capacitance. When trapping $^{40}Ca^+$ ion, a tunable capacitance is used to obtain a resonant frequency of about 24.5 MHz. In this case, assume the output voltage, resonant frequency, quality factor, input power and respectively are $V_1$, $f_1$, $Q_1$, $P_1$ respectively. Then connecting the high-voltage probe, these values are assumed to be $V_2$, $f_2$, $Q_2$, $P_2$ respectively. For the case of with the high-voltage probe and without the tunable capacitance, these values are supposed to be $V_3$, $f_3$, $Q_3$, $P_3$ respectively. Where, the output voltage $V_2$, $V_3$ can be measured by oscilloscope and the resonant frequency $f_1$, $f_2$, $f_3$ can be read out from function generator. And in all cases, the input power $P_1$, $P_2$, $P_3$ are 5 W respectively. Then from the equation (3), we can obtain the value of $(Q_2/Q_3)^{1/2}$. According to Ref. [18], it can be considered that the quality factor Q has a linear relationship with the resonant frequency f approximately when the resonant frequency has a very small change. Thus, $(Q_1/Q_2)^{1/2}$ can be determined from the value of $(Q_2/Q_3)^{1/2}$. Subsequently, the value of $V_1$ can be obtained according to the equation (3). Finally, based on the calculation described above, the output voltage of the trapping system and the dummy trap are determined to be 700 V and 520 V respectively. And the uncertainty of these two output voltage is evaluated to be 44 V and 48 V respectively. In modelling, the output voltage uncertainty is considered to be 50 V conservatively.

It can be clearly seen that the temperature of ring and endcap electrodes made from molybdenum is higher than other trap structure from the modelling results as well as the thermal picture. Moreover, the surfaces of these electrodes are visible to $^{40}Ca^+$ ion. Consequently, the uncertainty of molybdenum emissivity has a large contribution to the uncertainty of the radiation temperature to the ion. According to Ref. [13], molybdenum emissivity has been identified to be 0.16. In our work, the emissivity was estimated by adjusting thermal camera measurement of molybdenum according to the temperature of black tape on it. And we found these two values of emissivity could coincide with each other well. For polished molybdenum, the surface emissivity changes from 0.05~0.1 to 0.25 due to the difference surface oxidation. Therefore, an uncertainty of 0.09 is chosen for molybdenum emissivity in modelling.

RF absorption in ceramics has been confirmed to be one of the major heating sources in the trap. The power of heating due to RF absorption has relation with relative permittivity and dielectric loss tangent according to the following equation [13]:

$$p = 2\pi f \varepsilon_0 \varepsilon_r \tan(\delta) |E|^2, \tag{4}$$

where f is the frequency of the electric field, $\varepsilon_0$ is the permittivity of free space, $\varepsilon_r$ is the real relative permittivity and tan ($\delta$) is the dielectric loss tangent. Dielectric loss tangent of alumina is strongly dependent on its composition, processing and purity ranges from $1.0 \times 10^{-4}$ to $2 \times 10^{-4}$. Thus, $1.5 \times 10^{-4}$ and $0.5 \times 10^{-4}$ is adopted as dielectric loss tangent of alumina and its uncertainty in evaluating. Moreover, the lower thermal conductivity of ceramics can influence heat dissipation of ring electrode and has a large contribution on thermal distribution of the trap structure. For commercial alumina ceramics, thermal conductivity ranges from 20 to 35 W m$^{-1}$ K$^{-1}$. So, the thermal conductivity and its uncertainty is estimated to be 27.5 W m$^{-1}$ K$^{-1}$ and 7.5 W m$^{-1}$ K$^{-1}$, respectively.

The temperature of the room and the interior space of vacuum chamber have been monitored for more than 48 hours. The room temperature fluctuates less than 2 K around 292.20 K, while the one in vacuum chamber has a fluctuation less than 0.20 K. As the electrodes are separated from the feedthrough by ceramics, their temperatures can be considered to have same homogeneity with the room temperature. So the minimum and maximum initial temperature of the electrodes can be estimated to be 291.20 K and 293.20 K conservatively. For the ion confined in the vacuum chamber, it can be considered to possess a fluctuant temperature coinciding with the one in the interior space of vacuum chamber. Namely, it has a contribution of 0.10 K to the uncertainty of the radiation temperature seen by ion.

3.3 Evaluation of temperature rise seen by the $^{40}$Ca$^+$ ion

Figure 4 shows modelled temperature distribution of the miniature Paul trap with a 700 V RF at 24.5 MHz. It can be seen that the temperature of the ring electrode together with the middlemost titanium wire is highest, which is approximate 306 K. And the temperature of ring electrode and endcap electrodes is determined to be 305.39 K and 304.39 K respectively. From the simulating process, it can be clearly seen that the alumina insulator connected to the ring electrode is firstly heated up. That may mainly attribute to RF absorption in insulators, as well as joule heating in conductors, which two have been identified as the two main sources of heat generation in the ion traps [13]. Therefore, it can reduce the temperature rise of the ring electrode as well as other electrodes by adopting insulators with lower dielectric loss tangent and metal electrodes with higher electrical conductivity.

A small ball with black body properties has been placed at the ion position to evaluating the radiation temperature and its uncertainty seen by $^{40}$Ca$^+$ ion. The feasibility of this have been discussed in the paper [13]. According to whether or not have relation with RF heating, RF heating and fluctuation of room temperature have be taken into consideration to evaluate the uncertainty of temperature rise respectively, and the evaluation results are listed in table 3. It can be seen that temperature rise seen by the ion caused by RF heating is about 1.72 K and the two largest contributions to the uncertainty are coming from emissivity of molybdenum and RF amplitude, which are 0.42 K and 0.13 K respectively. Meanwhile, the contribution of room temperature fluctuation can't be neglected. Due to fluctuating with the room temperature, initial temperature of electrodes have been regarded to be 291.20 K and 293.20 K respectively and has a contribution of 0.06 K to the uncertainty of radiation temperature. Moreover, the temperature fluctuation in the interior space of vacuum chamber has been identified to possess a contribution of 0.10 K to the uncertainty of the radiation temperature seen by ion. All in all, the radiation temperature rise seen by $^{40}Ca^+$ ion in the miniature Paul trap is evaluated to be 1.72 K with an uncertainty of 0.46 K. According to Ref. [11], the temperature associated contribution to uncertainty of BBR shift is 2.2 mHz, corresponding to a fractional uncertainty $5.4 \times 10^{-18}$. While, the total uncertainty of BBR shift is 5.3 mHz and the uncertainty of BBR shift coefficient provides a major contribution of 4.8 mHz, which is at $10^{-17}$ level in fractional units. Obviously, the temperature associated uncertainty of BBR shift is much smaller than the one caused by the BBR shift coefficient.

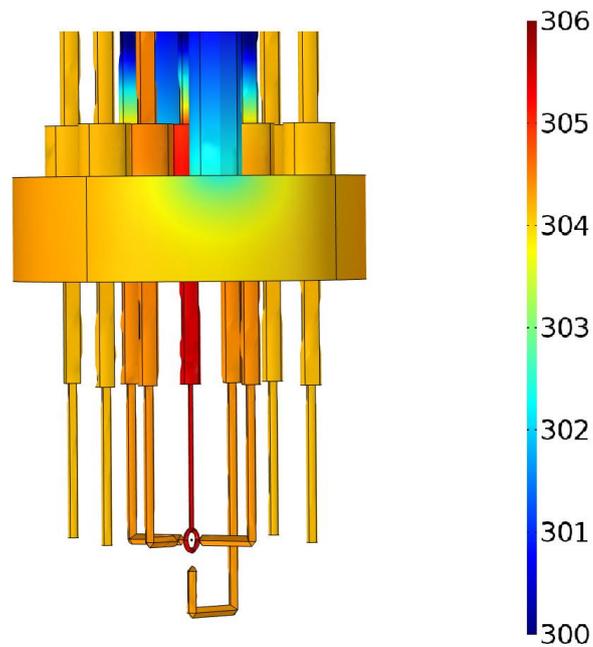

**Figure 4.** Modelled temperature distribution of the miniature Paul trap with a 700 V RF at 24.5 MHz.

**Table 3.** Evaluation of BBR temperature rise seen by $^{40}Ca^+$ ion in the miniature trap.

| Source | | Temperature rise seen by ion | Uncertainty |
|---|---|---|---|
| RF heating | RF amplitude | 1.72 | 0.13 |
| | Emissivity of Mo | | 0.42 |
| | Thermal conductivity of alumina | | 0.01 |
| | Dielectric loss tangent of alumina | | 0.01 |
| Fluctuation of room temperature | Initial temperature of electrodes | 0 | 0.06 |
| | Vacuum chamber temperature | | 0.10 |
| Total | | 1.72 | 0.46 |

4 Conclusion

As most of ion optical clocks, BBR shift has been one of the largest contributions to the systematic uncertainty of $^{40}Ca^+$ ion optical clock. BBR temperature rise seen by $^{40}Ca^+$ ion and its uncertainty was evaluated via FEM modelling. The FEM model was validated by comparing with the trap temperature at several points on the dummy trap measured by thermal camera. The measurements of thermal camera were calibrated by using PT1000 resistance thermometer for obtaining the actual temperature of molybdenum in the vacuum chamber. The uncertainty of input modelling parameters and their contributions to the uncertainty of the radiation temperature rise were evaluated in detail. The evaluating result shows that the radiation temperature rise seen by $^{40}Ca^+$ ion in the miniature Paul trap is 1.72 K with an uncertainty of 0.46 K. That means it has a contribution of 2.2 mHz and $5.4 \times 10^{-18}$ in fractional units to BBR shift uncertainty of $^{40}Ca^+$ ion optical clock. While, the total uncertainty of BBR shift is 5.3 mHz which is mainly contributed by the BBR shift coefficient with 4.8 mHz and at $10^{-17}$ level in fractional units.

Although, the evaluation has reduced BBR shift uncertainty due to temperature to 2.2 mHz. It could be further improved via optimizing the trap structure. According to the model, the heating is mainly from RF absorption in insulators and joule heating in conductors. Thus, it is essential to adopt insulators with lower dielectric loss tangent and metal electrodes with higher electrical conductivity to reduce the temperature rise of electrodes. Moreover, the radiation temperature rise seen by ion can be reduced using electrodes with lower emissivity and higher thermal conductivity. Simultaneously, more precise measurements and estimations on the modelling parameters are essential for reducing their contributions to the systematic uncertainty.

Acknowledgment

We would like to thank Huang Yao, Guan Hua, et al for their assistance on experiment and providing thermal camera. The work is supported by the National High Technology Research and Development Program of China (863 Program) (Grant No. 2012AA120701) and the National Natural Science Foundation of China (Grant No. 11174326).